\documentclass[aps,prd,twocolumn,eqsecnum,amssymb,amsmath,showpacs,a4paper, superscriptaddress]{revtex4-2}

\usepackage{graphicx}
\usepackage{amsfonts}
\usepackage{amsmath}
\usepackage{hyperref}
\usepackage{units}
\usepackage{color}
\usepackage{dcolumn}
\usepackage{bm}
\usepackage{float}
\usepackage{cleveref}
\usepackage[normalem]{ulem}
\usepackage{appendix}
\usepackage{mathtools}
\usepackage{esvect}

\usepackage[colorinlistoftodos, size=tiny, bordercolor=white]{todonotes}
\usepackage{comment}

\graphicspath{ {images/} }
\usepackage{mathrsfs}
\usepackage{amssymb}
\usepackage{dsfont}
\usepackage{enumitem}
\usepackage{gensymb}
\usepackage{bm}

\usepackage{mathtools}
\usepackage{chngcntr}
\counterwithout{equation}{section}

\makeatletter
\let\cat@comma@active\@empty
\makeatother

\newcommand{\grad}{\bm{\nabla}}

\newcommand{\tp}{{tp}}

\begin{document}

\title{Quantum vortex stability in draining fluid flows}

\author{Sam~Patrick}
\email{samuel_christian.patrick@kcl.ac.uk}
\affiliation{Department of Physics, King’s College London, University of London, Strand, London, WC2R 2LS, UK}

\date{\today}

\begin{abstract}
\noindent
Quantum vortices with more than a single circulation quantum are usually unstable and decay into clusters of smaller vortices. One way to prevent the decay is to place the vortex at the centre of a convergent (draining) fluid flow, which tends to force vortices together.
It is found that whilst the primary splitting instability is suppressed in this way (and completely quenched for strong enough flows) a secondary instability can emerge in circular trapping geometries.
This behaviour is related to an instability of rotating black holes when superradiantly amplified waves are confined inside a reflective cavity.
The end state of the secondary instability is dramatic, manifesting as a shock wave that propagates round the circular wall and nucleates many more vortices.
\end{abstract}
\maketitle

\section{Introduction}

Quantum vortices are topological defects in the order parameter $\Psi(\mathbf{x},t)$ describing a quantum fluid, e.g. Bose-Einstein condensates (BECs), polariton fluids, superfluid $^4$He, photon superfluids etc.
Their dynamics is of widespread interest due to their role in various important phenomena, including superfluid turbulence \cite{barenghi2001quantized}, topological phase transitions \cite{kosterlitz2017nobel} and superconductivity \cite{abrikosov2004nobel}.
The circulation of a quantum vortex is $\ell\kappa$, that is, $\ell$ units of circulation quantum $\kappa = h/M$, with $h$ Planck's constant and $M$ the mass of (bosonic) particles in the fluid.
The integer $\ell$ counts the number of times the phase $\mathrm{arg}(\Psi)$ winds around the defect in the order parameter.
The most basic quantum vortex has $|\ell|=1$ and is called a singly quantised vortex (SQV), whereas a vortex with $|\ell|>1$ is called a multiply-quantised vortex (MQV).

Since the energy of a vortex scales with $\ell^2$, MQVs tend to split into clusters of SQVs as a result of an energetic instability.
Hence, any mechanism that provides a reservoir to dissipate energy (e.g.\ coupling to a thermal bath) will cause an MQV to decay \cite{barenghi2016primer}.
Furthermore, even in ideal conservative systems, vortices can couple to sound waves (phonons) in the fluid to produce a dynamical instability \cite{shin2004dynamical,isoshima2007spontaneous}.
Because of this, many studies of quantum vortices typically employ configurations of SQVs. 

There are, however, certain conditions under which MQVs do not decay.
For example, energetic favorability of MQVs is found in rotating condensates with anharmonic (quartic) traps \cite{lundh2002multiply} or pinning potentials \cite{simula2002stability}.
Furthermore, the dynamical splitting instability can be suppressed in finite size systems which possess a discrete phonon spectrum \cite{giacomelli2020ergoregion}.
Another stabilisation mechanism involves filling the vortex core with a second condensate, such that there is an energy barrier associated with splitting the MQV \cite{richaud2023mass,patrick2023stability}.
A well known property of vortices (which follows from Kelvin's circulation theorem \cite{lautrup2011physics}) is that they tend to move under the influence of the local fluid velocity.
Hence, a natural stabilising mechanism involves placing an MQV at the focal point of a convergent fluid flow, e.g.\ by draining fluid from the system in a small region in the centre.
If the drain is strong enough, one would expect the convergence of the fluid to override the tendency of the MQV to split, rendering the vortex stable against fragmentation.

Convergent fluid flows in superfluid $^4$He were realised in \cite{yano2018observation,matsumura2019observation,obara2021vortex}
and the flow was reported to contain $\mathcal{O}(10^4)$ vortex quanta.
Motivated by these experiments, \cite{inui2020bathtub} simulated the evolution of vortex filaments around the drain at non-zero temperature, unveiling a formation mechanism for the vortex bundle above the drain.
3D simulations based on a Gross-Pitaevskii model containing a low number of vortices were performed in \cite{ruffenach2023superfluid}, which showed the convergent flow tends to focus vortices over the drain hole, twisting them into a bundle that funnels the flow into the drain below.
More recently, a novel method of surface wave spectroscopy was used to constrain the core size of a macroscopic draining vortex containing $\mathcal{O}(10^4)$ vortex quanta \cite{svanvcara2023exploring}.

MQV stability has also been demonstrated in exciton-polariton condensates, systems with inherent particle losses due to the finite lifetime of the polariton quasiparticles \cite{carusotto2013quantum}.
Simulations in \cite{alperin2021multiply} showed that a ring-shape laser beam induces particle fluxes toward the centre, capable of stabilising the MQV situated there.
Furthermore, wavevector dependent losses can induce a convergent flow centred on each vortex, leading to an attractive force between that causes vortices to merge \cite{solnyshkov2023towards}.
Stability of an $\ell=15$ vortex was recently demonstrated in the experimental set-up of \cite{delhom2023entanglement}, although in that case, the fluid velocity is forced by the pump laser and the convergent flow is weak since the polariton decay is not restricted to a central region.

In atomic BECs, convergent flows can be realised using an electron beam to ionise particles and deplete the condensate in a localised region.
Numerical simulations in \cite{zezyulin2014stationary} showed that an $\ell=2$ vortex can be stabilised by strong localised dissipation, and that the vortex splits once dissipation is switched off.
Giant vortex aggregates were reported in a rotating condensate in which a focused laser beam was used to remove atoms on the rotation axis \cite{engels2003observation}.
These vortex aggregates were attributed to the combined effect of the Coriolis force and the radial influx of atoms.

In light of these results, one may expect that MQVs are unambiguously stabilised under a strong enough convergent flow.
The purpose of this work is to demonstrate that this is not necessarily the case.
It is shown that even when the primary splitting instability is quenched, secondary instabilities capable of driving the system far from equilibrium can arise.
These secondary instabilities are related to the black hole bomb instability of Press and Teukolsky \cite{press1972floating}.
A rotating black hole is surrounded by an ergoregion, where fluctuations in a certain frequency range possess negative energies. 
Incoming radiation can scatter into these negative energy states, causing the escaping fraction of radiation to become amplified.
This process is called \textit{rotational superradiance} \cite{brito2020superradiance}.
The black hole bomb proposal involves encircling the system with a reflective mirror that scatters amplified radiation back into the black hole where it is further amplified, leading to a runaway process.
A quantum vortex, which from the perspective of low frequency sound waves can be viewed as an analogue rotating spacetime \cite{slatyer2005superradiant}, also possesses an ergosphere inside its core.
Indeed, the MQV splitting instability can be viewed as a negative energy bound state inside the ergosphere, which couples to sound waves outside the vortex to release energy \cite{giacomelli2020ergoregion,patrick2022origin,patrick2022quantum}.
Dissipation inside the vortex core leads to damping of this negative energy mode, akin the classical formulation of black hole superradiance where the negative energy is absorbed by the horizon \cite{cardoso2022dissipative}.
We will show that, when the vortex is located at the centred of a circularly symmetric trapping potential, superradiantly amplified sound can trigger the black hole bomb instability.

The structure of this paper is as follows.
In Sections~\ref{sec:setup} and \ref{sec:vort}, we describe the stationary draining vortex profiles of a quantum fluid.
In Section~\ref{sec:eigen}, we analyse the eigenmodes of the system for some illustrative parameters, demonstrating the effect of the convergent flow on stability.
Section~\ref{sec:wkb} provides an interpretation of the instabilities using a WKB approximation of the resonance formula for unstable eigenmodes.
In Section~\ref{sec:num}, numerical simulations of the fully nonlinear equations of motion are described, revealing the dramatic effect of the secondary instability on the late time dynamics.
Section \ref{sec:conc} concludes and implications of the results are discussed.

\section{Set-up} \label{sec:setup}

For concreteness, we consider a tightly confined atomic BEC with quasi-2D dynamics in the $\mathbf{x}=(x,y)$--plane.
In the mean field approximation, the system is governed by the Gross-Pitaevskii equation (GPE),
\begin{equation} \label{GPE}
    i\hbar\partial_t\Psi = \left[-\frac{\hbar^2 \nabla^2}{2M} +V(\mathbf{x}) - i\Gamma(\mathbf{x}) + g|\Psi|^2-\mu\right]\Psi,
\end{equation}
where $\Psi$ is the order parameter, $M$ is mass of the particles (which are bosons), $\mu$ is the chemical potential, $g$ is the 2D interaction parameter, $V(\mathbf{x})$ is a trapping potential and $\Gamma(\mathbf{x})$ is a dissipation term.
In the context of the BEC, localised dissipation can be achieved by illuminated the condensate with an ionising beam \cite{barontini2013controlling}.
The GPE applies equally to polariton condensates \cite{delhom2023entanglement}, photon superfluids \cite{braidotti2022measurement} and can even be considered as a phenomenological model of superfluid $^4$He on small scales \cite{volovik2003universe}.
We consider traps with a hard circular wall located  $r=r_B$. Specifically, we use the function,
\begin{equation}
    V = \frac{V_0}{1+(V_0-1)e^{a(r_B-r)}},
\end{equation}
where $V_0$ is the height of the trap and $a$ is a smoothing parameter. Unless specified otherwise, we mostly use $V_0=a=5$ and $r_B=25$ in the dimensionless units defined below in \eqref{adim}.
In the Madelung representation, $\Psi=\sqrt{n}e^{iM\Phi/\hbar}$ where $n(\mathbf{x})$ is the particle number density and $\Phi(\mathbf{x})$ a phase whose gradient gives the velocity field $\mathbf{v}=\grad\Phi$.
The GPE then separates into a Bernoulli equation,
\begin{equation} \label{Bern}
    \partial_t\Phi + \frac{1}{2}\mathbf{v}^2 + \frac{gn+V(\mathbf{x})-\mu}{M} = \frac{\hbar^2}{2M^2}\frac{\nabla^2\sqrt{n}}{\sqrt{n}},
\end{equation}
where the term on the right is the quantum pressure modification and a continuity equation,
\begin{equation} \label{cont}
    \partial_t n + \grad\cdot\left(n\mathbf{v}\right) = -2\Gamma(\mathbf{x})n.
\end{equation}
From the last equation, it is apparent that the function $\Gamma$ acts a sink for the density.
We consider a dissipation function localised at the centre of the trap of the form,
\begin{equation}
    \Gamma = \Gamma_0 \frac{1+\tanh[\kappa(r_0-r)]}{2},
\end{equation}
where $\Gamma_0>0$ is the strength of dissipation, $r_0$ is the radius of region where particles are lost and $\kappa$ is a smoothing parameter which we set equal to $2$ in the units of \eqref{adim}.
When $\Gamma_0=0$, the GPE conserves the total number of particles in the condensate $N=\int d^2\mathbf{x}\,n$.
For finite $\Gamma_0$, $N$ decreases according to,
\begin{equation} \label{dNdt}
    \partial_t N = -2\int d^2\mathbf{x}\,\Gamma n.
\end{equation}
Particles then flow inward to replenish the depleted condensate inside the dissipation region, establishing a convergent fluid flow.
This is the quantum analogue of the classical draining bathtub flow that forms over a physical fluid outlet \cite{andersen2003anatomy,stepanyants2008stationary} and the dissipation region can be viewed at the analogue of the drain hole.
In order to establish a stationary state in the simulations of the following sections, particles will be resupplied near the edge of the trap to hold $N$ fixed.

\section{Vortices} \label{sec:vort}

We search for stationary states containing vortices at the centre of the trap.
The condensate phase and velocity field are,
\begin{equation} \label{DBT}
    \Phi = \frac{\hbar}{M}\left[\ell\theta + \Theta(r)\right], \quad \mathbf{v} = \frac{\hbar}{M}\left[\frac{\ell}{r}\hat{\mathbf{e}}_\theta + \partial_r \Theta \hat{\mathbf{e}}_r\right],
\end{equation}
where $\mathbf{v}$ has the profile of a draining vortex.
The radial part of the phase $\Theta(r)$ is an unknown function that must be solved for. Note that our assumption of a circular trap with the draining vortex located at the centred implies the stationary background state described by $(n,\mathbf{v})$ is a function of $r$ only.

We start by considering the non-draining case when $\Gamma_0=0$ and $v_r\equiv\mathbf{e}_r\cdot\mathbf{v}=0$. 
In that case, the stationary solutions are the typical quantum vortices, whose velocity field is fully determined by the constraint of an irrotational fluid flow and the density $n$ is the only function one has to solve for.
The size of the vortex core is set by the healing length,
\begin{equation}
    \xi_0 = \hbar/\sqrt{M\mu_0},
\end{equation}
where $\mu_0=\mu(\Gamma_0=0)$. We can then simplify the calculation by defining the following dimensionless quantities,
\begin{equation} \label{adim}
\begin{split}
    \frac{\mathbf{x}}{\xi_0}\to & \ \mathbf{x}, \quad \frac{\mu_0t}{\hbar}\to t, \quad \frac{\mu}{\mu_0}\to\mu, \\
    & \frac{gn}{\mu_0}\to n, \quad \frac{M\Phi}{\hbar}\to\Phi.
\end{split}
\end{equation}
In these units, the chemical potential for the non-draining vortex is unity by definition.
In equilibrium, the density is determined by the equation,
\begin{equation}
    \frac{1}{2}\nabla_r^2 z + \left[1-\frac{\ell^2}{2r^2}-V(r)\right]z = z^3,
\end{equation}
where $z = \sqrt{n}$ and $\nabla^2_r  = \frac{1}{r}\partial_r r\partial_r$.

The presence of a radial component of $\mathbf{v}$ complicates the flow pattern. To simplify the expressions, we work with the dimensionless quantities defined in \eqref{adim}.
In equilibrium, \eqref{cont} implies $\partial_r(rnv_r) = -2r\Gamma n$ which is solved by,
\begin{equation} \label{vr1}
\begin{split}
    v_r(r) = & \ -\frac{2\int_0^{r} dr' r'\Gamma(r') n(r')}{n(r)r}, \\
    = & \ \begin{cases}
        -D/r, \qquad \qquad \, r\ll r_0 \\
        -\Gamma_0 r/(\ell+1), \quad r\gg r_0,
    \end{cases}
\end{split}
\end{equation}
where $D = 2\int dr\,r\Gamma n/\mu$ is the drain constant and the integral is performed over the region where $\Gamma(r)$ has support.
Hence, approaching the centre from the boundary, the radial velocity magnitude increases as the fluid converges on the dissipation region.
Inside the dissipation region, particles are lost and mass conservation causes a decrease in the radial velocity magnitude, with $v_r$ vanishing at the very centre.
Using these properties of $v_r$, \eqref{Bern} implies the following asymptotics for the density,
\begin{equation} \label{n_limits}
\begin{split}
    n(r) = & \ \begin{cases}
        \mu - (\ell^2+D^2)/2r^2, \qquad \quad r\gg r_0 \\
        r^{2\ell}\left[A + B r^2 +\mathcal{O}(r^4)\right], \quad r\ll r_0
    \end{cases} \\
B = & \ A[A\delta_{\ell 0}-\mu]/(\ell+1),
\end{split}
\end{equation}
where the value of $A$ can only be determined from the full solution and $\delta_{\ell 0}$ is the Kronecker delta function.
For $\ell=0$, $A>0$ since it is the value of the density at $r=0$. 
It will turn out that $A<\mu$ so that $B<0$, i.e.\ $n$ has negative curvature at $r=0$.
This implies that the density is a non-monotonic function of $r$.
When $\ell\neq 0$, $A$ is also positive since it determines the rate at which the density rises from zero at the vortex centre.

To search for equilibrium states when $\Gamma_0\neq 0$, we apply the following prescription.
Starting from the non-draining solution, we switch on a finite value of $\Gamma_0$ so that the density in the region $r<r_0$ begins to deplete.
We then resupply the depleted particles near the edge of the trap. The over density near $r_B$ will cause the condensate to flow inward toward the centre, establishing a convergent radial flow.
The fixed particle number $N$ is accommodated by a shift in the chemical potential.
The details of this procedure are described in Appendix~\ref{app:stat}.

\begin{figure}
\centering
\includegraphics[width=\linewidth]{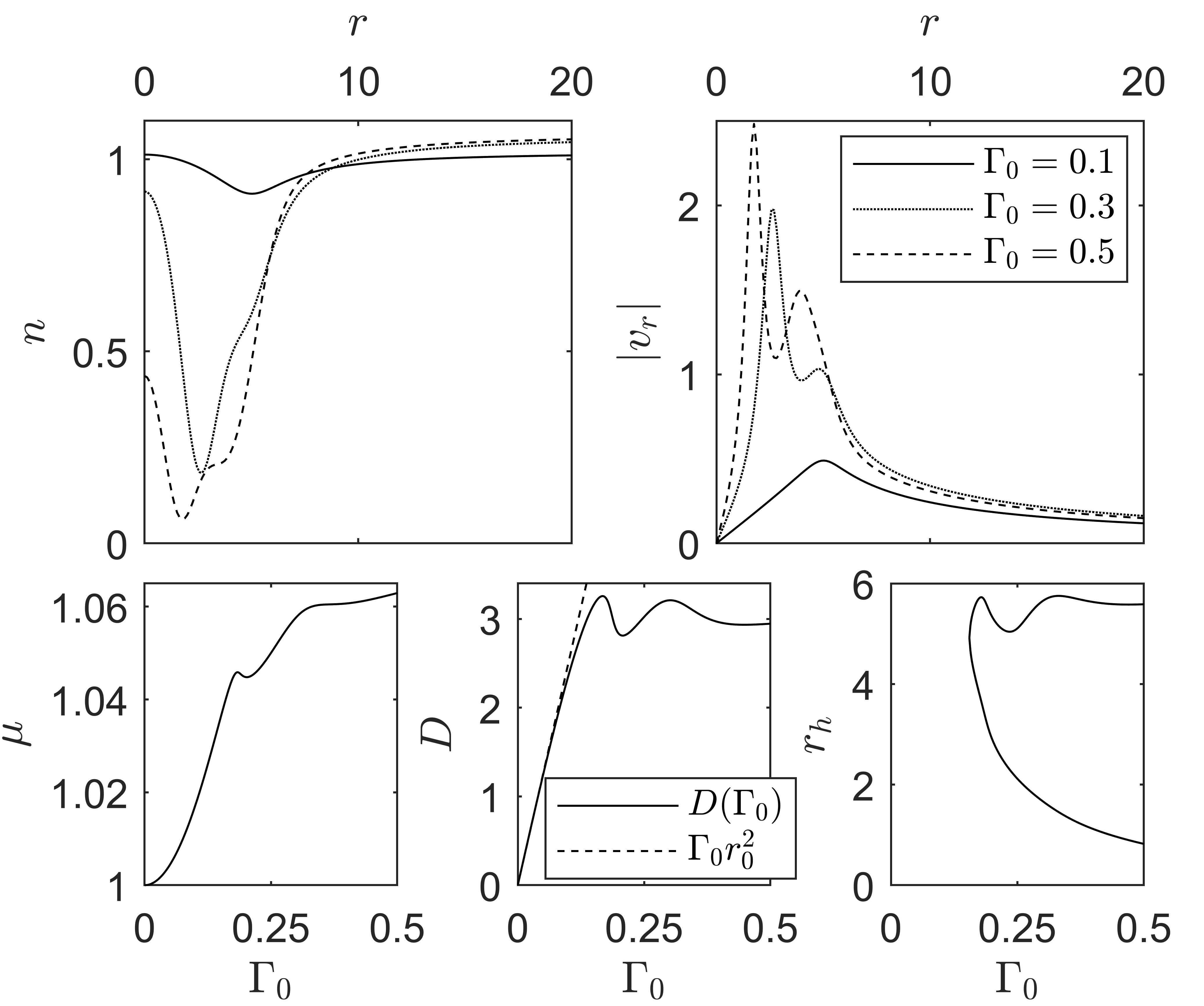}
\caption{Stationary draining flow profiles for $r_0=5$. In the top panels, we display the density $n$ and radial velocity magnitude $|v_r|$ for three illustrative values of $\Gamma_0$. On the lower panels, we show the variation of the chemical potential $\mu$, the drain rate $D$ and the horizon radii $r_h$ with $\Gamma_0$.} \label{fig:1}
\end{figure}

In Fig.~\ref{fig:1}, we display some examples of the density and radial velocity profiles when there is no vortex present. The density exhibits a depression moving toward the centre followed by a sudden rise at the very centre. This is the behaviour anticipated in \eqref{n_limits}.
The magnitude of the radial velocity displays the properties anticipated in \eqref{vr1}, although the behaviour in the intermediate region can be intricate due to the non-monotonic nature of the density.
As expected, the chemical potential adjusts as we increase $\Gamma_0$ but keep $N$ fixed.
The general increase in $\mu$ results from $v_r$ lowering the density in the dissipation region, thereby pushing atoms into the bulk and raising the asymptotic density level.
However, the variation of $D(\Gamma_0)$ reveals a curious feature that beyond a certain $\Gamma_0$, stronger dissipation can weaken radial flow, i.e.\ the more one tries to drain the fluid, the less one manages to achieve it.
This is related to the macroscopic Zeno effect studied in \cite{zezyulin2014stationary}.
We also display the locations $r_h$ satisfying $(n-v_r^2)|_{r=r_h}=0$.
In the region where $n$ is close to its asymptotic value, this represents the location where radial flow exceeds the sound speed $\sqrt{n}$, i.e.\ the acoustic horizon \cite{basak2003superresonance}.
Note that, since $v_r$ decreases toward the origin, the horizon necessarily comes as a black hole-white hole horizon pair.

\begin{figure}
\centering
\includegraphics[width=\linewidth]{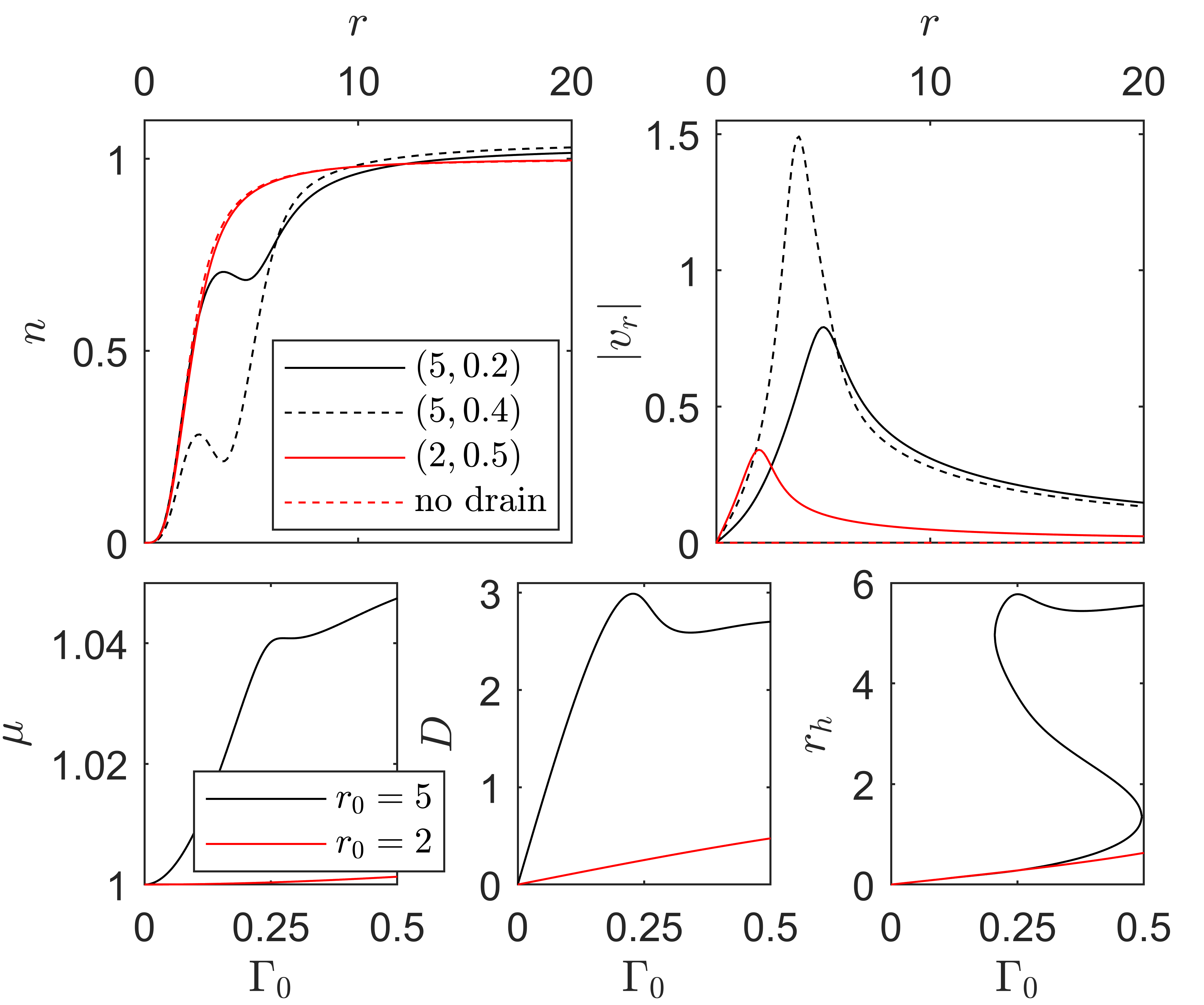}
\caption{In the top two panels, we show the density and radial velocity profiles when there is an $\ell=2$ located at $r=0$.
The legend gives the dissipation function parameters in the format $(r_0,\Gamma_0)$.
The radial velocities are much weaker than the $\ell=0$ case since the vortex reduces the density in the dissipation region.
In the lower panels, we show the dependence of $\mu$, $D$ and $r_h$ on the dissipation strength for two different values of $r_0$. 
Decreasing the size of the dissipation region weakens the draining flow.} \label{fig:2}
\end{figure}

In Fig.~\ref{fig:2}, we illustrate the effect of placing a vortex (in this case one with $\ell=2$) in the centre of the dissipation region. Since the density drops to zero on the vortex axis, there are fewer particles present in the dissipation region to be damped, hence, the radial velocity is smaller when compared with the $\ell=0$ case.
The black lines represent $r_0=5$ (same as Fig.~\ref{fig:1}) whereas the red lines are for $r_0=2$, illustrating that the draining flow is weaker when the dissipation region is smaller as one would expect.

\section{Eigenmodes} \label{sec:eigen}

To find out whether the stationary profiles obtained above are stable, we analyse their perturbations.
The perturbed order parameter may be written,
\begin{equation}
    \Psi = e^{i\Phi}\left[\sqrt{n} + \sum_{m\geq0} \left(u e^{im\theta} + v^*e^{-im\theta}\right)\right],
\end{equation}
where $u=u(r,t;m)$ and $v=v(r,t;m)$ are the radial eigenfunctions.
Under this decomposition, the linearised equations may be written as,
\begin{equation} \label{BdG}
\begin{split}
    & i\partial_t\begin{pmatrix}
        u \\ v
    \end{pmatrix} = \begin{bmatrix}
         \mathcal{D}_+ - i\widetilde{\Gamma} & n \\ -n & -\mathcal{D}_- - i\widetilde{\Gamma}
    \end{bmatrix} \begin{pmatrix}
        u \\ v
    \end{pmatrix}, \\
& \mathcal{D}_\pm = -\frac{1}{2}\nabla^2_r +\frac{(\ell\pm m)^2}{2r^2} +\frac{v_r^2}{2} \mp iv_r\partial_r + 2n + V - \mu, \\
& \widetilde{\Gamma} = \Gamma + \grad\cdot\mathbf{v}/2.
\end{split}
\end{equation}
This implies,
\begin{equation} \label{norm}
    \partial_t \mathcal{N} = -2\int d^2\mathbf{x} \, \Gamma\rho_n,
\end{equation}
where $\mathcal{N}=\int d^2\mathbf{x}\, \rho_n$ is the norm and \mbox{$\rho_n=|u|^2-|v|^2$} is the norm density, i.e.\ the norm isn't conserved in the presence of a dissipative mechanism.
Unlike the previous section where we resupplied $N$ near $r_B$, we do not replenish the dissipated mode norm since our reinjection strategy in Section~\ref{sec:num} will assume all lost particles are resupplied to the axisymmetric background solution.
Notice that the norm density (and hence the norm) can be either positive or negative.

\begin{figure*}
\centering
\includegraphics[width=\linewidth]{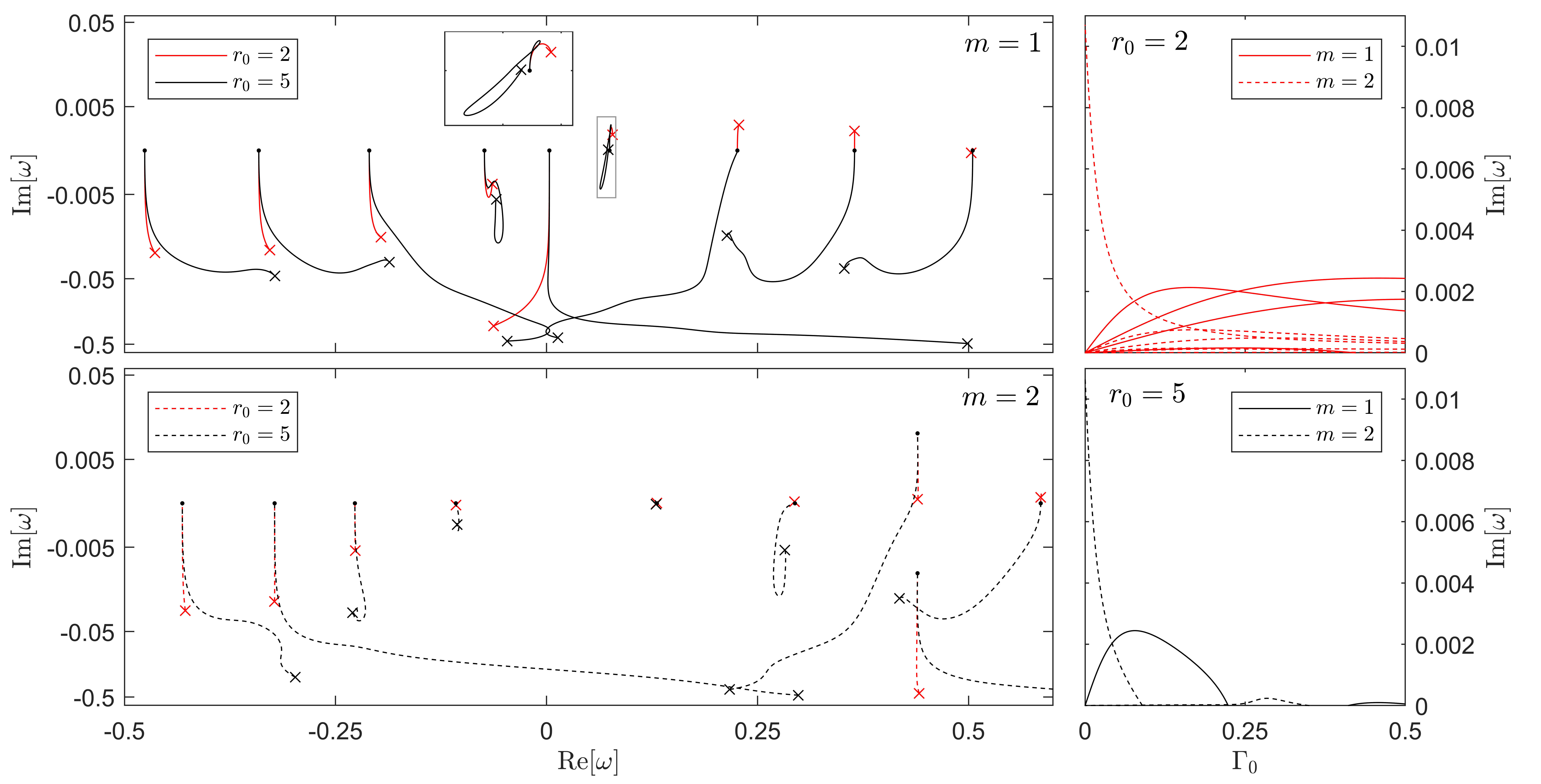}
\caption{\textbf{Left panels:} Eigenvalue trajectories through the complex plane as $\Gamma_0$ is varied. The trajectories start at the solid dots ($\Gamma_0=0$) and end on crosses ($\Gamma_0=0.5$). We show results for $m=1,2$ and $r_0=2,5$. In the top panel, the inset shows in detail the trajectories of the modes inside the grey rectangle. Note the nonuniform spacing on the vertical axis. \textbf{Right panels:} Unstable modes for $r_0=2$ (top) and $r_0=5$ bottom. The conventions for the line style and color correspond to those on the left panels.
For most modes in the spectrum (including the MQV instability) the localised dissipation tends to make the mode decay. However, certain modes attain a positive imaginary part when $\Gamma_0>0$, representing a new class of unstable modes. Fewer secondary instabilities are found for larger $r_0$.
} \label{fig:3}
\end{figure*}

In the stationary state, the operator on the right hand side of \eqref{BdG} is $t$-independent and the equation is separable. Writing $(u,v)^\mathrm{T} = a(t)(\tilde{u},\tilde{v})^\mathrm{T}$, one finds,
\begin{equation} \label{BdG2}
    \omega\begin{pmatrix}
        \tilde{u} \\ \tilde{v}
    \end{pmatrix} = \begin{bmatrix}
         \mathcal{D}_+ - i\widetilde{\Gamma} & n \\ -n & -\mathcal{D}_- - i\widetilde{\Gamma}
    \end{bmatrix} \begin{pmatrix}
        \tilde{u} \\ \tilde{v}
    \end{pmatrix},
\end{equation}
and $i\partial_t a = \omega a$, which is solved by $a(t)=a(0)e^{-i\omega t}$ where $\omega=\omega^r+i\omega^i\in\mathbb{C}$.
The equation for the norm in \eqref{norm} implies the modulus of $a$ behaves like,
\begin{equation} \label{imag_freq}
    |a(t)| = |a(0)|e^{\omega^it}, \qquad \omega^i = -\frac{1}{\mathcal{N}} \int d^2\mathbf{x} \,\Gamma \rho_n,
\end{equation}
where $\omega^i$ is a spatial averaging of the dissipation function weighted by the norm density.
For uniform dissipation, we find $\omega^i=-\Gamma$ and all modes are damped irrespective of the sign of their norm.
When $\Gamma$ varies in space, the sign of $\omega^i$ depends on the sign of the norm density in the dissipation region.
In that case, both growing (i.e.\ unstable) and decaying modes can exist.

Note that the naive expectation that the dissipation function should damp positive energy modes and enhance negative energy ones (where \mbox{$\mathcal{H} = \mathrm{Re}[\omega]\mathcal{N}$} is the mode energy) is not satisfied here.
Unlike damping found in stochastic formulations of the GPE \cite{cockburn2009stochastic}, the dissipative term in \eqref{GPE} decreases the particle number rather than the energy, which is the statement of \eqref{dNdt}.
The important thing for modes is whether $\rho_n$ in the dissipation region has the opposite sign to $\mathcal{N}$, as expressed by \eqref{imag_freq}.

To solve \eqref{BdG2} for the eigenvalues $\omega$ and radial eigenfunctions $(\tilde{u},\tilde{v})^\mathrm{T}$, we diagonalise the matrix in square parentheses using standard numerical algorithms (see Appendix~\ref{app:eig}).
Fig.~\ref{fig:3} shows the eigenvalues $\omega$ obtained in this way for the $\ell=2$ vortex.
The left panels show the trajectories of the eigenvalues through the complex plane as $\Gamma_0$ is varied, for $m=1$ (top panel) and $m=2$ (bottom panel).
We also show the effect of varying the size of dissipation region by displaying results for $r_0=2$ (red lines) and $r_0=5$ (black lines).
As $\Gamma_0$ increases, most modes in the spectrum tend to migrate down into the lower half plane where $\mathrm{Im}[\omega]<0$. However, certain modes can acquire a small positive imaginary part when the dissipation is turned on, with more unstable modes arising for smaller dissipation regions.
This is further exemplified by the panels on the right, which displays the variation of the eigenvalues $\mathrm{Im}[\omega]>0$ with $\Gamma_0$.
The mode with $\mathrm{Im}[\omega]>0$ at $\Gamma_0=0$ is the MQV splitting instability, which stabilises as the dissipation is increased.
However, other modes in the spectrum can be unstable for $\Gamma_0\neq 0$, with the $m=1$ modes displaying the largest tendency for instability.

\section{WKB method} \label{sec:wkb}

There are three key features of the spectra in Fig.~\ref{fig:3} that require explanation: (1) the quenching of the MQV splitting instability and (2) the existence of secondary instabilities which are (3) fewer in number for larger $r_0$.
To explain this behaviour, we apply the WKB method of \cite{patrick2022quantum}, the application of which to our problem is detailed in Appendix~\ref{app:lin}.
The basis of the approximation is to write fluctuations as a superposition of in- and out-going plane waves with a radial wavenumber $p$ which varies with $r$ due to the inhomogeneous flow field.
Schematically these two modes of wave motion oscillate as,
\begin{equation} \label{WKB_modes}
    e^{i\int (p^\pm+i\Gamma\partial_\omega p^\pm) dr+im\theta-i\omega t},
\end{equation}
where $p^\pm$ are the two solutions to local dispersion relation, which asymptotically looks like the standard Bogoliubov one, i.e.\ \mbox{$(\omega-\mathbf{v}\cdot\mathbf{k})^2=nk^2+k^4/4$} with $k=|\mathbf{k}|$.
Note that this gets modified inside the vortex core (see Appendix~\ref{app:lin} for details).
Throughout the analysis, $\omega$ is treated as a real quantity and the imaginary part obtained at the end of the procedure as a next-to-leading order correction once the boundary conditions have been imposed.

\begin{figure}
\centering
\includegraphics[width=\linewidth]{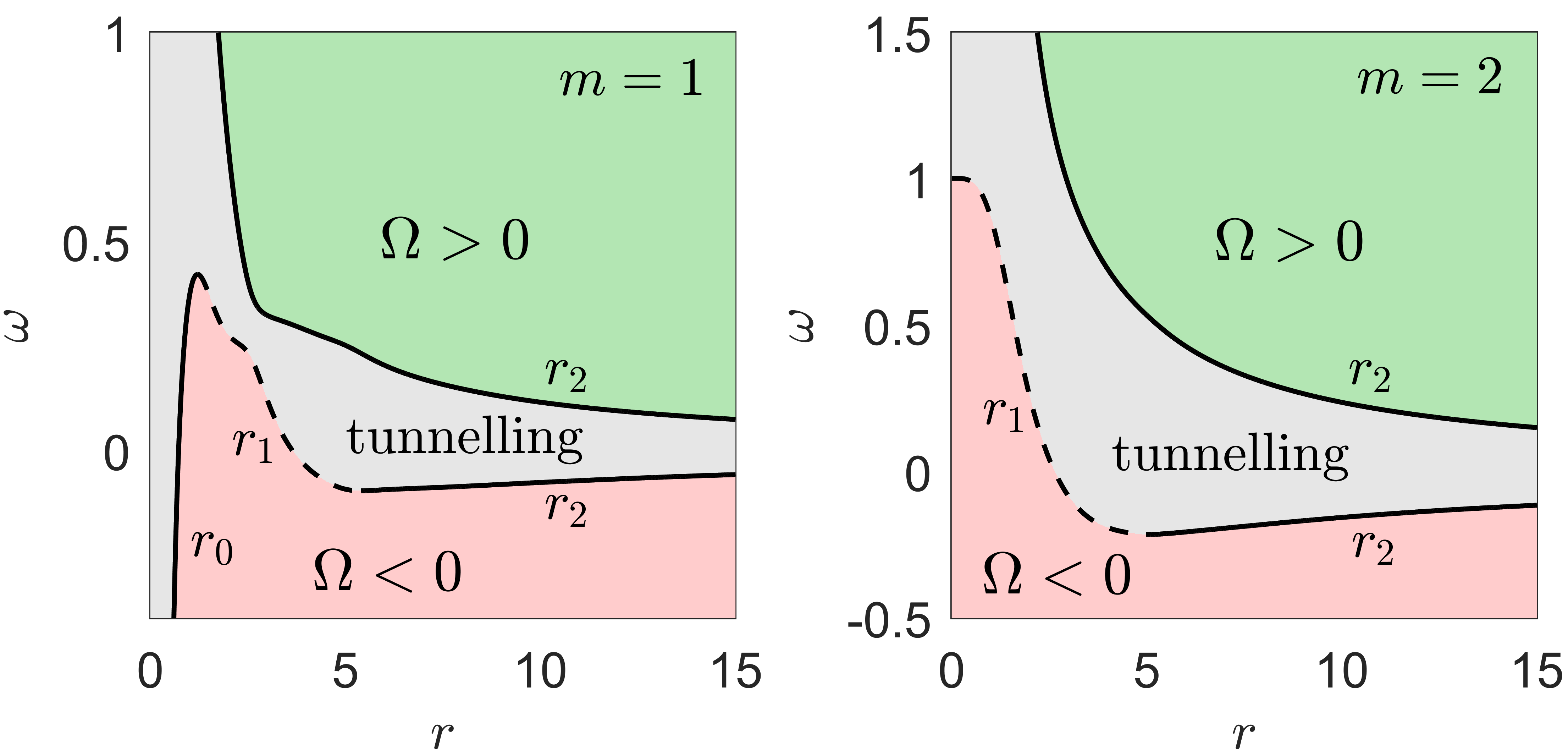}
\caption{The WKB modes in \eqref{WKB_modes} can exhibit three different behaviours: propagating with positive $\rho_n$ (green region), propagating with negative $\rho_n$ (red region) or tunnelling (grey region).
These regions are delineated by the turning points $r_{0,1,2}$.
The curves shown are for the parameters $\Gamma_0=0.05$, $r_0=5$, $\ell=2$ and $m=1,2$.
Note that for $m=\ell$, the inner turning point is $r_0=0$ by convention.} \label{fig:4}
\end{figure}

The term in the exponent of \eqref{WKB_modes} proportional to $\Gamma$ describes the decay of the wave amplitude in the direction of propagation.
The $e^{i\int p^\pm dr}$ has different behaviour in three separate regions: two propagating regions where $\Omega=\omega-\mathbf{v}\cdot\mathbf{k}$ is positive and negative respectively and a zone separating the two where the wave tunnels.
These regions are depicted on Fig.~\ref{fig:4}.
Since $\rho_n\propto\Omega$ in the WKB approximation, a mode with a given $\omega,m$ has a positive (negative) norm density when it is in the green (pink) area.
Since the energy density is proportional to $\mathrm{Re}[\omega]\rho_n$, a frequency that crosses from the green to the pink region undergoes superradiant amplification (see \cite{patrick2022origin,patrick2022quantum} for details on this point).

We can derive a resonance condition for modes which satisfy the boundary conditions.
For those $\omega$ which cross the tunnelling region in Fig.~\ref{fig:4}, it is given by,
\begin{equation} \label{cotcotexp}
    4\cot S_{01}\cot(S_{2B}+\pi/4) = e^{-2I_{12}},
\end{equation}
where we have defined,
\begin{equation} \label{phase_integral}
\begin{split}
    2S_{ab} = & \ \int^{r_b}_{r_a}\left(1+i\Gamma\partial_\omega\right)(p^+-p^-) dr, \\
    I_{12} = & \ \int^{r_2}_{r_1}|\mathrm{Im}[p^+]| dr.
\end{split}
\end{equation}
The derivation of this formula is given in Appendix~\ref{app:lin}.
The $S$ integrals can be split into real and imaginary parts $S_{ab}=S^r_{ab}+iS^i_{ab}$, where these two terms are respectively the phase integral from $r_b$ to $r_a$ then back again, and the total damping accumulated along the trajectory.
When the frequency needed to solve this condition has a small imaginary part, we make the replacement \mbox{$S^r_{ab}\to S^r_{ab}+i\omega^i\partial_\omega S^r_{ab}$},
whilst $S^i_{ab}$ is evaluated for $\omega^r$ since $\Gamma_0$ is assumed small.
The $I$ integral is evaluated in a region where $p^\pm$ are complex conjugate pairs, and the effect of damping has been ignored in computing it since its contribution to \eqref{cotcotexp} is already exponentially small.

Before delving into the novel features of the spectrum when $\Gamma_0>0$, we first we show how \eqref{cotcotexp} encodes the physics of the basic MQV splitting instability in the case $\Gamma_0=0$.
This condition describes two types of eigenmode in the system.
The phonon frequencies (i.e.\ modes which live outside the vortex) are exponentially close to the zeros of \mbox{$\cot (S_{2B}+\pi/4)$}. Similarly, excitations of the vortex core have frequencies close to the zeros of $\cot S_{01}$. 
However, when the zeros of the two cotangent functions are close, the exponential on the right of \eqref{cotcotexp} leads to a coupling between the two types of eigenmode.
Let $\cot S_{01}(\omega_v)=0$ and \mbox{$\cot [S_{2B}(\omega_p)+\pi/4]=0$}.
In that case, we expand \eqref{cotcotexp} around the zeros,
\begin{equation}
    (\omega-\omega_p)(\omega-\omega_v) + \epsilon |\mathcal{T_{\bar{\omega}}}|^2/T_pT_v \simeq 0,
\end{equation}
where $|\mathcal{T}_{\bar{\omega}}|\simeq e^{-I_{12}(\bar{\omega})}$ is the modulus of the transmission coefficient across the tunnelling zone evaluated at the central frequency \mbox{$\bar{\omega}=(\omega_p+\omega_v)/2$}, $\epsilon = \mathrm{sgn}(S'_{01}S'_{2B})$ with the prime denoting an $\omega$ derivative, and \mbox{$T_p = 2|S'_{2B}|$} and \mbox{$T_v=2|S'_{01}|$} are the crossing times for phonons and vortex modes.
In our case, $\epsilon=+1$ for $\omega>0$ (which superradiate) and $\epsilon=-1$ for $\omega<0$ (which do not superradiate).
The eigenfrequencies are then approximately given by,
\begin{equation}
    \omega \simeq \bar{\omega} \pm \sqrt{\Delta\omega^2 -\epsilon \gamma^2}.
\end{equation}
Hence for non-superradiant states, there is an avoided crossing whereas for superradiant states, we obtain a complex conjugate pair (one of which is unstable) when the mode coupling $\gamma=|\mathcal{T}_{\bar{\omega}}|/\sqrt{T_pT_v}$ is larger than the frequency splitting $\Delta\omega=(\omega_p-\omega_v)/2$. 
This coupling of the negative energy vortex mode to a positive energy phonon is precisely the reason for the dynamical splitting instability, as the MQV radiates energy into sound when it splits (see \cite{patrick2022origin,patrick2022quantum} for a comparison between the exact spectrum and the predictions of \eqref{cotcotexp}).

For the discussion of $\Gamma_0\neq 0$, let us rewrite \eqref{cotcotexp} in the two following equivalent forms,
\begin{align}
    X +e^{2iS_{01}} + ie^{2iS_{2B}}\left(1+Xe^{2iS_{01}}\right) = & \ 0, \label{res_vort} \\
    1 + iXe^{2iS_{2B}} + e^{-2iS_{01}}\left(X+ie^{2iS_{2B}}\right) = & \ 0, \label{res_phon}
\end{align}
where we have defined,
\begin{equation}
    X = \frac{1+e^{-2I_{12}}/4}{1-e^{-2I_{12}}/4},
\end{equation}
which is the modulus of the reflection coefficient for superradiant modes and the inverse of the same for non-superadiant modes.

Condition \eqref{res_vort} is amenable for studying the MQV splitting instability when the propagation time between the vortex and $r_B$ is sufficiently large that $|e^{2iS_{2B}}|\ll 1$ (e.g.\ for low frequencies the relevant condition is $r_B\gg1/\omega^i$).
Note that for this exponential to be small, we must have an unstable mode with $\omega^i>0$, which only occurs for $\omega^r>0$.
If we assume $\omega^r\gg\omega^i$, the oscillation frequency is determined to leading order by \mbox{$\cos S^r_{01}(\omega^r) = 0$}, describing a mode in the vortex core, and the growth rate by,
\begin{equation} \label{MQVopen}
    \omega^i = \frac{\log X -2|S^i_{01}|}{T_v}.
\end{equation}
The first term shows that these
modes are unstable due to superradiance, since $X>1$ and therefore the logarithm contributes positively.
The second term results from dissipation and acts to reduce the amplitude of the instability, which is consistent with Fig.~\ref{fig:3}. Physically this makes sense since the dissipation acts in the region where the mode lives, and from \eqref{norm} we expect dissipation to deplete the norm density (or by proxy the amplitude) rather than the energy.
Therefore, in this case, superradiance tries to make the mode grow whereas dissipation tries to stop it growing.
This explains the first feature of Fig.~\ref{fig:3} identified at the beginning of this section.

To understand the origin of the secondary instabilities, we study the second condition \eqref{res_phon}.
Consider $\omega^r>0$ modes in the limit where dissipation is sufficiently strong that $|e^{-2iS_{01}}|\ll 1$.
To satisfy this condition, the vortex core should be large enough that the mode propagating inside it has enough time to be damped significantly, which is achieved in practice for large $\ell$.
For a small growth rate, the leading order condition for the oscillation frequency is \mbox{$\cos(S^r_{2B}(\omega^r)+\pi/4)=0$} describing phonons trapped outside the vortex. The growth rate is given at leading order by,
\begin{equation} \label{BHB}
    \omega^i = \frac{\log X-2|S^i_{2B}|}{T_p},
\end{equation}
where the logarithmic term again contributes positively due to superradiance.
When dissipation is confined to the region where the mode propagates in the vortex core, $S^i_{2B}=0$ and phonons become unstable, explaining the second finding of Fig.~\ref{fig:3}.
Physically, the negative energy part of the mode responsible for superradiant amplification gets damped out inside the dissipation region and (to leading order) the part of the wave which reflects back can be neglected.
This is the same mechanism as the black hole bomb instability, where superradiant modes outside the black hole are confined by an external mirror and the horizon provides a perfectly absorbing inner boundary.
When there is dissipation in the green region of Fig.~\ref{fig:4}, $S^i_{2B}$ becomes non-zero and acts to weaken the instability, completely quenching it for large enough values.
This explains the third finding of Fig.~\ref{fig:3} that secondary instabilities are suppressed for large dissipation regions.
Note that although we have considered here two specific limits (first large $r_B$ then large $\ell$) we expect the same qualitative features
to emerge in the spectrum for general $r_B$ and $\ell$.

\begin{figure*}
\centering
\includegraphics[width=\linewidth]{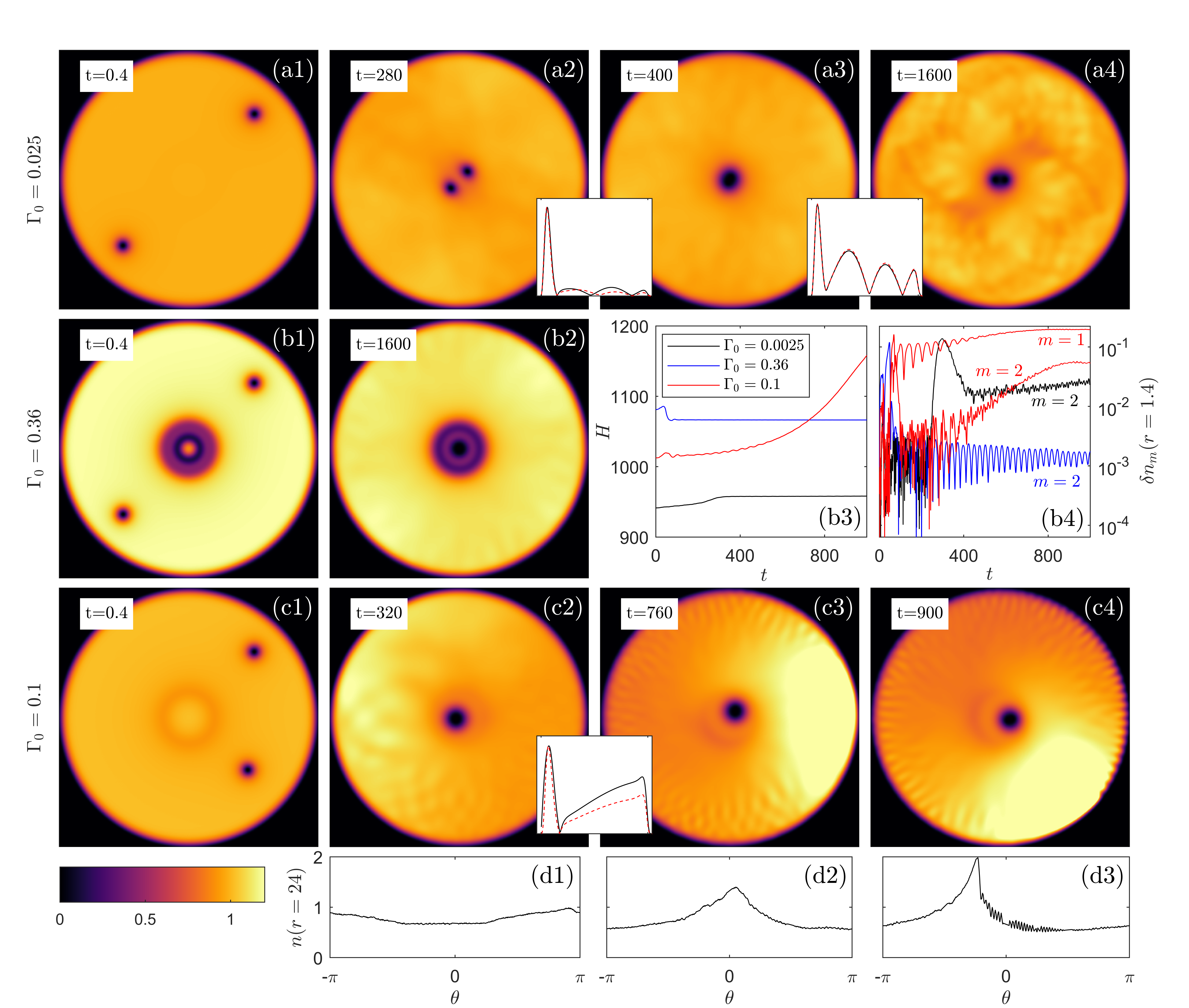}
\caption{Intensity plots of $n(x,y)$ from three different simulations, which differ by the dissipation strength and the initial vortex locations. We used $r_0=5$ and $r_B=25$ and the $x$ and $y$ limits for each panels are $[-r_B,r_B]$.
Panels (a1-4) show a low $\Gamma_0$ simulation, and the insets ($r\in[0,r_B]$ on the horizontal and a.u. on the vertical axis) show that the $m=2$ waveforms during vortex recombination and splitting are in good agreement with the corresponding eigenmodes in the spectrum.
Panels (b1-2) show that for higher dissipation, the vortices do not separate at late times.
Panels (c1-4) show that for an asymmetric initial condition, the inspiral can trigger a secondary instability in the $m=1$ mode.
The inset compares the measured waveform with corresponding $m=1$ eigenmode.
Panels (d1-3) show the shape of the unstable mode near the boundary in the angular direction.
At late times, the wave becomes sharply peaked, triggering a short wavelength instability at the leading edge.
Panels (b3-4) show the total energy and the most highly occupied $m$-components near the centre of the trap as functions of $t$.
} \label{fig:5}
\end{figure*}

\section{Numerical simulations} \label{sec:num}

In this section, we explore the predictions of Fig.~\ref{fig:3} by simulating the full nonlinear equation in \eqref{GPE}, demonstrating the effect of the convergent flow on a pair of initially separate vortices.
We take as our initial condition a stationary flow with $\Gamma_0\neq0$ but no vortex.
We then imprint a pair of SQVs (both rotating counter-clockwise) near the edge of the trapping potential.
Particles lost due to dissipation are replenished at the edge of the trap such that $N$ is conserved.
Details of the procedure are described in Appendix~\ref{app:num}.
Fig.~\ref{fig:5} shows the results of three such simulations.

\begin{figure*}
\centering
\includegraphics[width=\linewidth]{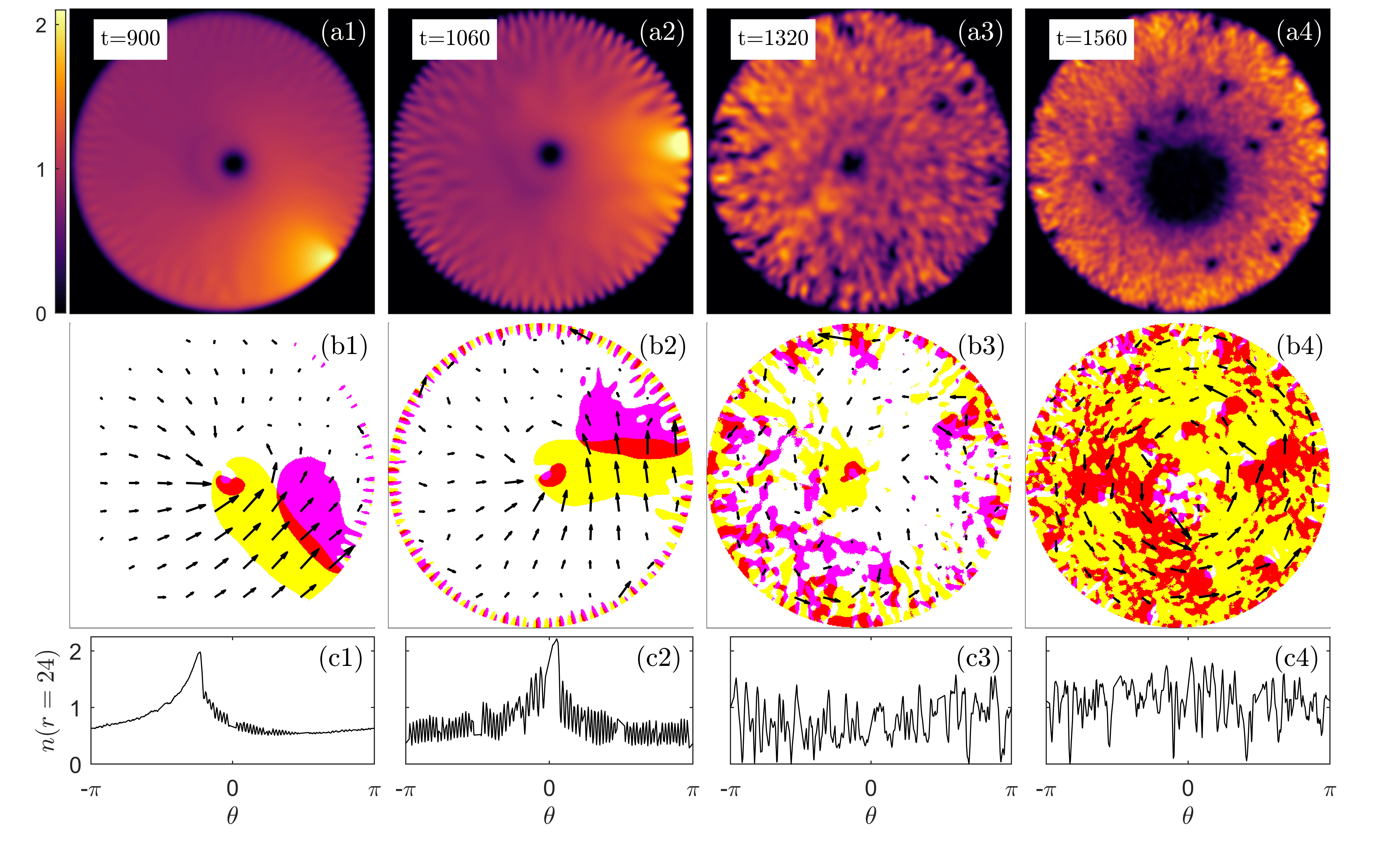}
\caption{Panels (a1-4) depict a continuation of panel (c4) in Fig.~\ref{fig:5} with the colour axis increased to clearly show the overdense region. 
Panels (b1-4) show a vector map of the velocity field at the same times, with regions where $v_r>0.06$ ($v_\theta>0.45$) highlighted in magenta (yellow).
On panels (c1-4), we see a shock wave being generated at the leading edge of the $m=1$ wave eventually propagating all the way round the trap edge.
Once the shock wave reaches large enough amplitudes, corotating vortices nucleate near the boundary (panel (a3)) and move into the centre to form a giant vortex mass (panel (a4)).
} \label{fig:6}
\end{figure*}

In panels (a1-4) of Fig.~\ref{fig:5}, the amount of dissipation is small and we start with the vortices located symmetrically about the centre at $(x,y) = (12.7,12.7)$ and the antipodal point.
The vortices quickly spiral into the middle and, once they are separated by around three healing lengths, the motion is well described by the conjugate mode to the MQV splitting instability, which describes vortex recombination.
Between panels (a2) and (a3), we Fourier transform $n_{m=2}$ over the interval $t=[317.2,410]$ and plot the $\omega$ component with the largest amplitude (black curve), which agrees well with the recombination mode predicted from \eqref{BdG2} (dashed red line).
The frequency is computed by performing a quadratic interpolation around the peak value of the Fourier transform at $r=1.4$ (where the amplitude of the spatial waveform exhibits a maximum).
The growth rate is found by performing a linear regression on \mbox{$\log|n_{m=2}(t,r=1.4)|$} over the same time interval.
We obtain a measured eigenfrequency $0.40(3)-0.0206(4)i$ \footnote{For the measured eigenfrequencies, the error in the real part is half the frequency resolution, whilst the error in the imaginary part is quoted from the $95\%$ confidence bounds. The predicted values are quoted to 3 s.f.}, which we compare against $0.439-0.0293i$ from Fig.~\ref{fig:3}.
The agreement is acceptable given that at this stage, the system is in a highly non-equilibrium situation.
Indeed, since $N$ is fixed, the value of $\mu$ will shift to accommodate the non-zero occupation of various excited states, leading to differences with the spectrum of Fig.~\ref{fig:2} which was computed in equilibrium. 
After a long time, the vortices separate by a small amount since the dissipation is not sufficient to quench the MQV splitting instability. The measured eigenfrequency of splitting mode (obtained using the same method as above over the interval $t\in[1000,1400]$) is \mbox{$0.44(1)+0.00150(3)i$} whereas the predicted value is \mbox{$0.440+0.00365i$}.
The poor agreement in the imaginary part can be attributed to the effects of $N$ conservation and nonlinearities for high mode amplitudes.
We checked this by performing a simulation starting from equilibrium with a small initial amplitude in the splitting mode, finding that, once the mode grew large enough, the growth rate dropped to around the measured value above.
The black curves on panels (b3-4) are the total energy $H$, which is a slowly increasing function at late times, and the $m=2$ mode amplitude which, after an initial transient, decreases then increases exponentially as expected from the discussion above.

In panels (b1-2) of Fig.~\ref{fig:5}, we show results of a simulation prepared with the vortices at the same locations but a larger value of $\Gamma_0$.
The two SQVs combine much quicker in this case and residual fluctuations dissipate away, in keeping with the fact that the spectrum contains no unstable modes for $m=1,2$.
The energy (blue curve on panel (b3)) decreases very slowly at late times whilst $|n_{m=2}|$ decreases exponentially, exhibiting a beating effect due to the superposition of two decaying co- and counter-rotating modes.
These results support the idea that a dense quantum vortex cluster can be prevented for splitting by a sufficiently strong dissipation mechanism.

In the third row of Fig.~\ref{fig:5}, we pick a configuration where the strongest unstable mode in the spectrum is one of the secondary instabilities with $m=1$.
Two SQVs are placed asymmetrically around the centre at $(x,y)=(12.7,12.7)$ and $(11.5,-10.2)$ to promote the excitation of this mode when they spiral into the middle and combine.
Between the second and third panels, we display the $m=1$ component of the density (which is the maximum of the time Fourier transform over the interval $t\in[500,800]$) and find good agreement with the predicted waveform.
The measured eigenfrequency of this mode is \mbox{$0.083(11) + 0.00068(1)i$} whilst the predicted value is \mbox{$0.0767 + 0.00236i$}.
The real part of the frequency agrees within the estimated error.
The discrepancy in the imaginary part is again attributed to the large amplitude during the measurement window (see Appendix~\ref{app:mode} for a discussion of this point).
Measurement of $\mathrm{Im}[\omega]$ at earlier times where the amplitude is smaller is complicated by the presence of a decaying counter-rotating mode, which causes the beating effect in the (red) $m=1$ curve on panel (b4).
Panel (b3) shows that there is a drastic increase in the total energy, which is due to the damping of the negative energy part of the wave in the dissipation region.
Panels (d1-3) illustrate the shape of the growing $m=1$ mode in the $\theta$ direction near the edge of the trap.
At late times, the wavefront becomes sharply peaked, which can be attributed to the nonlinear excitation of the $m=2$ wave (also shown on panel (b4)) which has roughly twice the frequency of the $m=1$ component.
In the last panel, a short wavelength instability (a \textit{shock wave}) appears at the leading edge of the overdense region and propagates around the trap wall. The observed self-steepening dynamics followed by the appearance of a shock wave is similar to that studied in the 1D set-up of \cite{damski2004formation}.

Further development of this state is depicted in detail in Fig.~\ref{fig:6}.
We show the density (panels (a)), the velocity field (panels (b)) and the angular profile of the density near $r_B$ (panels (c)) at four different instants of time.
Eventually, the angular shock wave becomes large enough that additional vortices nucleate at the trap wall, before being dragged into the centre by the draining flow, which is most apparent from panels (a).
The vector maps in the (b) panels represent the velocity field (with the arrows scaled with respect to the largest one in the plot).
We highlight regions where $v_r>0.06$ ($v_\theta>0.45$) in magenta (yellow).
The overlapping regions appear as red.
We find that the radial velocity is directed outward in front of the $m=1$ wave.
This supplies additional particles to the leading edge of the overdense region resulting in a sharp peak that continues to grow.
The angular velocity of the condensate is largest in the overdense region and decreases suddenly at the leading edge, in a manner consistent with known shock wave formation mechanisms \cite{damski2004formation}.
By the end of the simulation, the mass of vortices in the centre was precessing in the counter-clockwise direction whilst more SQVs continued to nucleate from the trap edge.

For the presented data in Fig.~\ref{fig:5} panels (c,d) and Fig.~\ref{fig:6}, the trap height was increased to $V_0=10^3$ to ensure the shock wave near the trap edge was not a consequence of the mirror image condensates arising from the periodic boundary conditions.
Qualitatively similar results were observed to those performed with $V_0=5$.
As a further check that the shock wave was not a consequence of the numerical resolution, we checked the effect of (firstly) doubling the space outside the $2r_B\times 2r_B$ condensate zone whilst keeping the same number of grid points, and (secondly) changing the number of grid points to $N_x=256$ (i.e. half the spatial resolution) and the time step to $\Delta t = 10^{-2}$.
The shock wave was found to form in both cases, with minor difference expected due to the change in the precise seeding conditions at different resolutions.





\section{Discussion} \label{sec:conc}

Multiply quantised vortices (MQVs) are generally expected to decay into clusters of singly quantised vortices (SQVs).
A natural mechanism purported to enhance the stability of MQVs involves a dissipation mechanism which induces a convergent (draining) flow in the system \cite{zezyulin2014stationary,alperin2021multiply,solnyshkov2023towards,svanvcara2023exploring}.
We have demonstrated that, in circularly symmetric trapping geometries (realisable experimentally \cite{johnstone2019evolution,navon2021quantum}) stability is not guaranteed, challenging the perspective that a convergent fluid flow has a stabilising influence on MQVs.

We considered a particle dissipation term used to model losses in Bose gases illuminated with a laser/electron beam \cite{engels2003observation,barontini2013controlling,zezyulin2014stationary}, polariton condensates \cite{carusotto2013quantum,alperin2021multiply,delhom2023entanglement} and nonlinear optical condensates (the so called \textit{photon superfluids}) \cite{zezyulin2012macroscopic,braidotti2022measurement}.
The model also captures phenomenological aspects of superfluid $^4$He flows in the thin film regime, with the density acting as a proxy for the height field and the dissipation representing loss of fluid from the plane, e.g.\ due to a physical outlet.

The observed tendencies of our system are a consequence of the dissipation reducing the particle number rather than the total energy which, when combined with an amplification mechanism, can result in secondary instabilities.
Our WKB method revealed that the relevant factor is where the confined mode lives relative to the localised dissipation region.
Since the MQV splitting instability is confined to the vortex core, where dissipation acts, it can be quenched.
When superradiant phonon modes are confined outside the dissipation region, the negative energy transmitted into the vortex core is absorbed whilst the mode occupation outside the vortex increases.
However, if the dissipation region extends outside the vortex core, the occupation of phonon modes can similarly be reduced.
We then demonstrated that secondary instabilities can have drastic implications for the late time dynamics of the systems, growing so large as to generate a shock wave at the edge of the condensate which seeds the nucleation of many vortices near the trap wall.

Observation of these features is within current experimental capabilities of existing platforms implementing atomic BECs.
For example, in \cite{gauthier2019giant}, a 2D planar geometry containing vortices was realised in a $^{87}$Rb condensate with a lifetime of $28\mathrm{s}$. 
In this system, the healing length is on the order of 1{\textmu}m and the natural timescale is $\tau = \xi/c \sim 1\mathrm{ms}$, where $c$ is the sound speed.
If dissipative losses are induced with an electron beam (e.g.\ \cite{zezyulin2014stationary,barontini2013controlling}) our results suggest that the secondary instability would onset on a timescale of $10^2-10^3\tau$, well within the condensate lifetime.

For polariton condensates, stability of an MQV with 15 circulation quanta was found in \cite{delhom2023entanglement}.
In that case, dissipation was uniform through the system, hence, superradiant phonons are dissipated as well as the MQV splitting mode.
The system size and phonon decay length were $\sim 70$ and $\sim 8$ healing lengths respectively, explaining the absence of secondary instabilities.
In our dimensionless units, their dissipative parameter is $\Gamma\sim0.1$, for which the WKB treatment of Section~\ref{sec:wkb} seems reasonable.
Hence, our methods could be applied to provide insight into properties of superradiance in these systems.

Finally, since Gross-Pitaevskii based models are a useful phenomenological tool for studying the microscopic dynamics of quantum liquids \cite{barenghi2001quantized}, our results may help elucidate the behaviour of superfluid vortex experiments.
In superfluid $^4$He, the core of a macroscopic vortex cluster is either a depression in the liquid's surface over the drain hole (which provides the effective dissipation) or a fully formed throat which plunges through it \cite{obara2021vortex,svanvcara2023exploring}.
Studies of the problem in 3D reveal the formation of a complex tangle of vortex lines over the drain \cite{inui2020bathtub,ruffenach2023superfluid}.
A detailed understanding of the simpler 2D problem may help identify the key mechanisms at play, and may be of further use in studying the effectively 2D dynamics of waves on the superfluid interface. \\

\textbf{Acknowledgements.}--- 
I would like to thank Maxime Jacquet and Ruth Gregory for helpful discussions.
I am also grateful for the hospitality of Perimeter Institute where part of this work was carried out.
This work was supported by the Science and Technology Facilities Council through the UKRI Quantum Technologies for Fundamental Physics Programme [Grant ST/T005858/1].
Research at Perimeter Institute is supported in part by the Government of Canada through the Department of Innovation, Science and Economic Development and by the Province of Ontario through the Ministry of Colleges and Universities.


\bibliographystyle{apsrev4-2}
\bibliography{main.bbl}

\appendix
\numberwithin{equation}{section}
\section{Numerical methods} 

\subsection{Stationary profiles} \label{app:stat}

First, we describe our numerical technique to find the stationary draining vortex profiles.
We exploit the symmetry of the desired solution by factoring out the vortex winding $e^{i\ell\theta}$ so that the equation to solve is a function of $r$ only.
Using the units of \eqref{adim}, we write \eqref{GPE} in the form,
\begin{equation} \label{GPEr}
    i\partial_t z = \left(-\frac{\partial_r^2}{2}-\frac{\partial_r}{2r} +\frac{\ell^2}{2r^2} + V-i\Gamma +|z|^2\right)z,
\end{equation}
where $z(r) = \sqrt{n(r)}e^{i\Theta(r)-i\mu t}$.
The radial grid is \mbox{$r_j \in [\Delta r,r_B+d]$}, and $j=1...N_r$ with $N_r=256$ and $d=5$.
For the radial derivatives, we use 5-point finite different stencils accurate to $\mathcal{O}(\Delta r^4)$, where $\Delta r$ is the radial grid step size.
The behaviour in \eqref{n_limits} implies that correct boundary conditions at $r=0$ are Neumann for $\ell=0$ and Dirichlet for $\ell\neq 0$.
The boundary condition at the edge of the grid is arbitrary since it is applied deep in the region where $z$ goes to zero under the potential barrier.
We choose it to be a Dirichlet boundary condition.

Due to the dissipation in the centre, any excitations of the ground state decay away. 
An implicit assumption here is that there are no unstable modes with $m=0$, which turns out to be true (at least for the scenarios we are interested in).
Therefore, starting from an arbitrary initial state, we should eventually arrive at a $z$ corresponding to the ground state provided we evolve \eqref{GPEr} for long enough.
The evolution is enacted using a fourth order Runge-Kutta algorithm.
$\Delta t = 2\times 10^{-3}$ is taken for the time step, satisfying $\Delta t<\Delta r^2/2$ which ensured stability.
This meant that going to higher spatial resolution was challenging since decreasing the spatial grid step size by a factor of 2 would increase the simulation time for the stationary states by a factor of 8, whilst the 2D simulations described in Appendix~\ref{app:num} would take 16 times longer.
After each time step, we compute the number of particles $\Delta N$ lost due to dissipation.
These particles are then resupplied at the edge of the trap before the next time step according to the following prescription,
\begin{equation}
\begin{split}
   & z\sqrt{1+\Delta N \beta(r)/|z|^2}\to z, \\
   & \beta = \beta_0\begin{cases}
       r-r_1, \quad \, r_1<r\leq r_2 \\
       1, \ \qquad \quad r_2<r\leq r_3 \\
r_B-r, \quad r_3<r\leq r_B \\
   \end{cases}
\end{split}
\end{equation}
where \mbox{$r_1 = r_B-4$}, \mbox{$r_2=r_B-3$} and \mbox{$r_3 = r_B-1$} and $\beta_0$ is a normalisation such that $2\pi \int dr\, r\beta = 1$. This ensures that the total number of particles is conserved throughout the simulation.

We estimate the chemical potential at each time step $t_k$ using $\mu_k = -\mathrm{Im}[\log(z_k/z_{k-1})]/\Delta t$ which gives a function over the $r$ grid.
The evolution is continued until the maximum value of $\mu_k-\mu_{k-1}$ in the region $r<r_B$ is below $5\times 10^{-6}$.
At this point there will still be residual fluctuations in $z$.
Since these are decaying, we evolve for a further 2000 ($10^4$) time steps for $\Gamma_0>0.05$ ($\Gamma_0\leq0.05$) and take the value of $|z|$ as its average during this time.
The phase is measured at each point $r_j$ and fit with the function $\Theta_j-\mu_j t$.
The first term in this fit gives the phase function in \eqref{DBT} from which obtain $v_r$ by enacting a $\partial_r$ stencil incorporating the Neumann boundary condition at the origin.
The second term gives the chemical potential as a function of $r$.
We observe that this is constant to a high precision and take its average over the region $r<r_B$.

To generate the solutions in Figs.~\ref{fig:1} and \ref{fig:2}, we start by calculating the non-draining solution with $\ell\neq 0$ using an imaginary time prescription (see e.g.\ Appendix~A of \cite{patrick2022origin}).
This is the solution for $\Gamma_0=0$.
We then use this as our initial condition for the largest value of $\Gamma_0$ simulated (which was $0.5$).
This was done since the strong dissipation here will quickly damp the solution to the desired equilibrium state.
We then generate the solutions for all $\Gamma_0$ in descending order, each time taking the result from the previous computation as the initial condition.

\subsection{Eigenmodes} \label{app:eig}

To solve \eqref{BdG2} for the eigenmodes, we use the same radial grid $r_j$ and finite difference stencils defined in the previous section.
We choose a Dirichlet boundary condition for the edge of $r$ grid, although the results are insensitive to this choice in the frequency range probed.
The boundary condition at $r=0$ is more delicate.
By expanding $(u,v)$ around $r=0$ and inserting into \eqref{BdG2}, we deduce that the leading order behaviour is $u\sim r^{|m+\ell|}$ and $v\sim r^{|m-\ell|}$.
Hence, we apply a Dirichlet boundary condition whenever the exponent is different from zero.
Since the next term in the series is multiplied by an extra factor of $r^2$, we apply a Neumann boundary condition for a zero exponent.
\eqref{BdG2} is then solved using a standard matrix diagonalisation routine (we apply Matlab's \verb|eig| function).

\subsection{Nonlinear dynamics in 2D} \label{app:num}

For the simulations in Section~\ref{sec:num} of the full equation \eqref{GPE}, we take as our initial condition a stationary solution with $\Gamma_0\neq0$ and no vortex, say $\Psi_0$, and interpolate this onto a Cartesian $(x,y)$ grid.
We take \mbox{$x_j\in[-(r_B+d),r_B+d]$} (and similarly for $y_j$) with $j=1...N_x$ and \mbox{$N_x=2N_r=512$}, so that the resolution is comparable to our radial code.
A pair of SQVs is then imprinted.
For this process, we first calculate the non-draining SQV density profile which asymptotes to unity, say $n_1(r)$, then extend the solution to a high value of $r$ using the expected asymptotic behaviour.
This solution is recentred to the vortex coordinates $(x_\nu,y_\nu)$ and the initial condition becomes,
\begin{equation} \label{init}
\begin{split}
    & \Psi_\mathrm{init} = \Psi_0(x,y) \\ & \times \prod_\nu n_1(x-x_\nu,y-y_\nu) \frac{(x-x_\nu)+i\ell_\nu(y-y_\nu)}{|\mathbf{x}-\mathbf{x}_\nu|},
\end{split}
\end{equation}
where the last factor ensures the defects have the correct phase winding.
We take $\ell_\nu=1$ with $\nu=1,2$ and $\mathbf{x}_\nu$ given in the main text.
Since the vortices merge in the centre during the simulation, we multiply $\Psi_\mathrm{init}$ by a factor $(N/\int d^2\mathbf{x}|\Psi_\mathrm{init}|^2)^{1/2}$ so that the system has the same total number of particles as our equilibrium solutions in Fig.~\ref{fig:2} (which, we recall, have the same $N$ as a non-draining MQV with unit chemical potential by construction).
The time evolution is then performed using a split-step Fourier spectral method.
That is, given the solution at time $t_k$, we find $\Psi$ at $t_{k+1}$ according to,
\begin{equation}
\begin{split}
    \widetilde{\Psi}_1 = & \ e^{-i\Delta t(V-i\Gamma+|\Psi_k|^2)/2}\Psi_k, \\
     \widetilde{\Psi}_2 = & \ \mathcal{F}^{-1}\big[e^{-i\Delta t\mathbf{k}^2/2}\mathcal{F}\big[\widetilde{\Psi}_1\big]\big], \\
     \Psi_{k+1} = & \ e^{-i\Delta t(V-i\Gamma+|\widetilde{\Psi}_2|^2)/2}\widetilde{\Psi}_2,
\end{split}
\end{equation}
where $\mathcal{F}$ and $\mathcal{F}^{-1}$ denote the 2D Fourier transform and its inverse on the $(x,y)$ grid, with $\mathbf{k}$ coordinates in Fourier space, and we set $\Delta t = 4\times 10^{-3}$.
At the end of each time step, we resupply particles lost through dissipation at the edge of the trap with the same prescription used to find our stationary solutions in the radial simulations.
In this way, the total particle number is conserved throughout the evolution.
We can think of this as representing a scenario where a fixed number of particles remain in circulation.

\begin{figure}
\centering
\includegraphics[width=\linewidth]{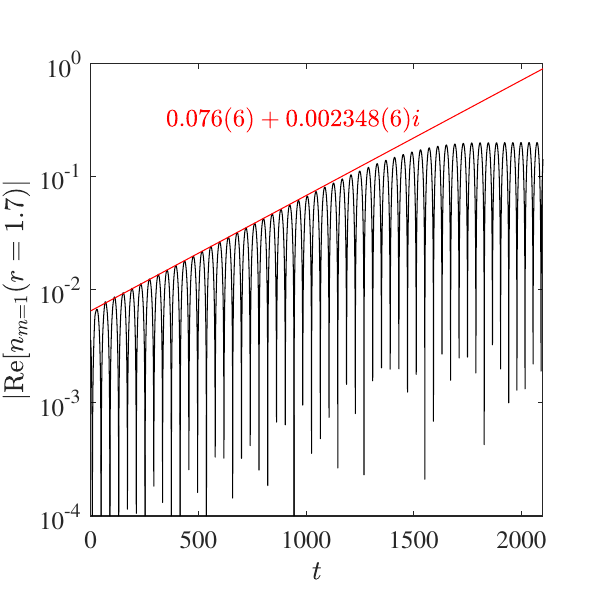}
\caption{Growth of the unstable $m=1$ mode (black line) with the same parameters used for Fig.~\ref{fig:5}. The best fit to the absolute value over the window $t\leq500$ is also shown (red line).
The measured eigenfrequency (displayed in figure) is in excellent agreement with the predicted value of \mbox{$0.0767 + 0.00236i$}.} \label{fig:7}
\end{figure}

\subsection{Mode growth} \label{app:mode}

In the simulations of Fig.~\ref{fig:5}, we found discrepancies between the measured eigenfrequencies and the predicted ones, particularly in the imaginary part.
We attribute this to nonlinear effects at large mode amplitudes as well as our resupply strategy, which keeps $N$ fixed at the expense of depleting the background for large mode occupations.
We illustrate an example of this on Fig.~\ref{fig:7} for the unstable $m=1$ mode.
The same parameters were used as for the (c) panels of Fig.~\ref{fig:5} except, this time, the initial condition was the stationary $l=2$ vortex state seeded with the (normalised) $m=1$ unstable mode multiplied by an amplitude $0.5$.
We measure the eigenfrequency for $t\leq 500$ and display the result on the figure.
The growth rate and oscillation frequency are in excellent agreement with the predicted values at early times when the amplitude is small.
At intermediate times, the growth rate starts to deviate from the initial value, with the amplitude saturating at late times. 
When the eigenfrequencies are measured in Fig.~\ref{fig:5}, the wave amplitudes are of the order $0.1$ which is where the discrepancy starts in Fig.~\ref{fig:7}.
This suggests that deviations in the growth rate are consistent with a nonlinear mechanism.

\section{WKB method} \label{app:lin}

In this appendix, we provide the details of the method leading to the resonance condition \eqref{cotcotexp} in the main text.
First, it is helpful to define new eigenfunctions $g,f$ according to,
\begin{equation}
    \sum_{m\geq 0} e^{im\theta}\begin{pmatrix}
        u \\ v
    \end{pmatrix} = \begin{bmatrix}
        g(\mathbf{x},t) + if(\mathbf{x},t)/2 \\
        g(\mathbf{x},t) - if(\mathbf{x},t)/2
    \end{bmatrix}, 
\end{equation}
which satisfy,
\begin{equation} \label{fg_eqn}
\begin{split} 
    (D_t-\widetilde{\Gamma})f + ng + \frac{1}{4}\mathcal{D}_r g = & \ 0, \\
    (D_t-\widetilde{\Gamma})g - \mathcal{D}_r f = & \ 0, \\
    \mathcal{D}_r = -\nabla^2 + \mathbf{v}^2 + 2(n+V- & \ \mu),
\end{split}
\end{equation}
and $D_t = \partial_t+\mathbf{v}\cdot\grad$ is the material derivative.
In the WKB approximation, we write,
\begin{equation}
    \begin{bmatrix}
        f \\ g
    \end{bmatrix} = \begin{bmatrix}
        \mathcal{A}(\mathbf{x},t) \\ \mathcal{B}(\mathbf{x},t)
    \end{bmatrix} e^{i\mathcal{S}(\mathbf{x},t)}.
\end{equation}
The amplitudes are assumed to vary gradually compared to the phase. 
That is, $|\partial\mathcal{A}|\ll|\mathcal{A}\partial \mathcal{S}|$ and $|\partial^2 \mathcal{S}|\ll(\partial \mathcal{S})^2$ (and similarly for $\mathcal{B}$) for $\mathbf{x}$ and $t$ derivatives.
Using the symmetry of the system, we can write,
\begin{equation}
    \mathcal{S} = \int p(r) dr + m\theta - \omega t \quad \begin{cases}
        \grad \mathcal{S} \equiv \mathbf{k} = (p,m/r), \\
        -\partial_t\mathcal{S} \equiv \omega
        \end{cases}
\end{equation}
where the decomposition into $\omega,m$ is exact since the background is independent of $t,\theta$.
Under these assumptions, \eqref{fg_eqn} becomes,
\begin{equation} \label{WKB_LO}
\begin{split}
    \Omega^2 = & \ n\widetilde{k}^2 + \widetilde{k}^4/4, \\
    \Omega = & \ \omega-\mathbf{v}\cdot\mathbf{k}, \qquad \widetilde{k}^2 = p^2 + \widetilde{m}^2/r^2, \\ \widetilde{m}^2 = & \ m^2+r^2\mathbf{v}^2+2r^2(n+V-\mu),
\end{split}
\end{equation}
at leading order and,
\begin{equation} \label{WKB_NLO}
\begin{split}
    & \partial_t\left(\frac{\Omega \mathcal{A}^2}{F}\right) + \grad\cdot\left(\frac{\mathbf{v}_g\Omega \mathcal{A}^2}{F}\right) = -\frac{2\Gamma\Omega\mathcal{A}^2}{F}, \\
    & F = n + \widetilde{k}^2/4,
\end{split}
\end{equation}
at next to leading order. 
A similar equation for the second amplitude can be derived using \mbox{$\mathcal{B} = i\Omega\mathcal{A}/F$}.
In the derivation of \eqref{WKB_LO} and \eqref{WKB_NLO}, we have assumed that dissipation is a weak effect that enters at second order. Hence, the method applies only for low $\Gamma_0$.

The first line of \eqref{WKB_LO} is the dispersion relation whose solutions $\mathbf{k}=\mathbf{k}(\omega,m;r)$ represent the local value of the wavevector for a given $\omega$ and $m$.
The group velocity of the wave is then determined by $\mathbf{v}_g = (\partial_\omega\mathbf{k})^{-1}$;
in particular, its radial component is $\hat{\mathbf{e}}_r\cdot\mathbf{v}_g = (\partial_\omega p)^{-1}$.
The equation in \eqref{WKB_NLO} describes how amplitude is transported, with the term on the right acting as a damping term.
Evaluating the norm density for a WKB mode gives $\rho_n = \Omega |\mathcal{A}|^2/F$ (where this expression is evaluated on the real part of $\omega$), so that \eqref{WKB_NLO} describes how the norm is transported across the system.
In particular, the sign of the norm density coincides with that of $\Omega$ and is transported in the direction of $\mathbf{v}_g$.
That is, a fluctuation with $\Omega<0$ has negative norm density.
The mode energy is defined as,
\begin{equation}
\begin{split}
    \mathcal{H} = & \ \int d^2\mathbf{x} \bigg[\frac{1}{2}|\partial_r u|^2 + \frac{1}{2}|\partial_r v|^2 + \frac{(\ell+m)^2}{2r^2}|u|^2 \\
    & \ -\frac{iv_r}{2}\left(u^*\partial_ru-u\partial_ru^*  - v^*\partial_rv + v\partial_rv^*\right), \\
    & \ + \frac{(\ell-m)^2}{2r^2}|v|^2 + (2n+V-\mu)(|u|^2+|v|^2)  \\
    & \ + n(uv^*+u^*v)\bigg].
\end{split}
\end{equation}
which we evaluate in the stationary regime with the aid of \eqref{BdG} to obtain,
\begin{equation}
    \mathcal{H} = \omega\mathcal{N} + i\int d^2\mathbf{x}\,\Gamma \rho_n. 
\end{equation}
From \eqref{imag_freq}, it then follows that $\mathcal{H}=\mathrm{Re}[\omega]\mathcal{N}$.
That is, a $\mathrm{Re}[\omega]>0$ wave with $\Omega<0$ (and therefore $\rho_n<0$) has negative energy density.

The dispersion relation \eqref{WKB_LO} has four distinct solutions representing the different allowed $\mathbf{k}$.
When $\Gamma_0$ is small, only two of these (which we call $\mathbf{k}^\pm$ such that $\mathrm{Re}[p^+]>\mathrm{Re}[p^-]$) correspond to propagating solutions.
The other two are evanescent everywhere and enforcing the boundary conditions sets their amplitudes to zero.
Hence, the general solution can be written as,
\begin{equation} \label{wave_sol}
\begin{split}
    f = & \ \sum_{\omega,m} e^{im\theta-i\omega t}\sum_{j=\pm} \frac{R^j(r)}{\sqrt{r}}, \\
    R^j = & \ \alpha^j\sqrt{\frac{F^j\partial_\omega p^j}{\Omega^j}} e^{i\int (p^j+i\Gamma\partial_\omega p^j) dr},
\end{split} 
\end{equation}
where $\alpha^j$ is an adiabatically conserved integration constant that comes from solving \eqref{WKB_NLO}, and superscript $j$ implies that the quantity is evaluated for one of the $p^\pm$ solutions to the dispersion relation. 
The integrals in the exponent are evaluated relative to a reference point where the phase is known.
Near $r_B$, $p^+$ corresponds to the out-going mode with \mbox{$\partial_\omega p^+>0$} whilst $p^-$ is in-going with \mbox{$\partial_\omega p^-<0$}.
Notice that the damping term in the exponent implies that the amplitude decays in the direction of the group velocity $(\partial_\omega p^j)^{-1}$ as one would expect.

The locations where the $p^j$ change from real (propagating) solutions to complex (evanescent) solutions are the turning points $r_\tp$. 
At these points, \mbox{$p^+_\tp = p^-_\tp$} and the group velocity in the radial direction vanishes, \mbox{$(\partial_\omega p^\pm_\tp)^{-1} = 0$}.
Intuitively, the wave propagating in one direction comes to rest instantaneously before propagating back in the other direction.
Hence, the turning points are the locations where waves scatter off the inhomogeneous background.
On the other side of $r_\tp$, $p^j\in\mathbb{C}$ and the waves tunnel.

Looking at \eqref{wave_sol}, we can see that the amplitude diverges as a consequence of the radial group velocity vanishing at $r_\tp$, signalling the breakdown of the WKB approximation.
This is not a problem but rather a necessity.
The WKB approximation encodes only the adiabatic variation of a given solution of the dispersion relation, i.e.\ the various solutions do not exchange energy with each other. To capture scattering (the transfer of energy between waves which travel in opposite directions) the WKB approximation must breakdown at certain locations.

There is a standard procedure to fix the breakdown of \eqref{wave_sol} which we now describe. The full equations of motion are expanded in the vicinity of $r_\tp$ and an exact solution for the scattering waves is obtained. The asymptotic form of this solution is matched onto the WKB modes either side of $r_\tp$ to provide a relation between them.
When waves propagate for $r<r_\tp$ and are evanescent for $r>r_\tp$, the relation is
\begin{equation} \label{tp1}
    \begin{pmatrix}
        R^+_\tp \\ R^-_\tp
    \end{pmatrix} = e^{\frac{i\pi}{4}}\begin{pmatrix}
        1 & -\frac{i}{2} \\ -i & \frac{1}{2}
    \end{pmatrix} \begin{pmatrix}
        R^\downarrow_\tp \\ R^\uparrow_\tp
    \end{pmatrix},
\end{equation}
whilst in the reverse scenario (evanescent for $r< r_\tp$ and propagating for $r>r_\tp$) we have,
\begin{equation} \label{tp2}
    \begin{pmatrix}
        R^\uparrow_\tp \\ R^\downarrow_\tp
    \end{pmatrix} = e^{\frac{i\pi}{4}}\begin{pmatrix}
        \frac{1}{2} & -\frac{i}{2} \\ -i & 1
    \end{pmatrix} \begin{pmatrix}
        R^+_\tp \\ R^-_\tp
    \end{pmatrix}.
\end{equation}
Here, a superscript $\uparrow$ ($\downarrow$) denotes a wave which grows (decays) in the direction of increasing $r$.

In Fig.~\ref{fig:4} of the main text, we depict the dependence of the various turning points $r_{0,1,2}$ on $\omega$ for $m=1,2$.
Between turning points, there is a region where the waves are evanescent and tunnel across the flow.
This tunnelling zone separates the region where \mbox{$\Omega>0$} \mbox{($\rho_n>0$)} from that with \mbox{$\Omega<0$} \mbox{($\rho_n<0$)}.
We then see that \mbox{$\mathrm{Re}[\omega]>0$} modes can have negative energy densities \mbox{($\mathrm{Re}[\omega]\rho_n<0$)} inside the vortex core, provided they are below a certain threshold (for $m=\ell$ this threshold is $\omega=\mu$).
Hence, when such a wave impinges on the vortex, it excites a state with negative energy in the core and (by energy conservation) gets amplified upon reflection.
This is the phenomenon of rotational superradiant scattering familiar from black hole physics \cite{brito2020superradiance}.

We can obtain a formula that describes the scattering of modes that cross this tunnelling zone in the following way.
Firstly, using \eqref{wave_sol}, we define a number that shifts one of the $R^j$ modes at $r_b$ to a smaller radius $r_a$, such that the WKB amplitude does not diverge for $r_a<r<r_b$.
That is, 
\begin{equation} \label{shift}
\begin{split}
    \mathcal{F}^j_{ab} = & \ \sqrt{\frac{\Omega^j_b F^j_a\partial_\omega p^j_a}{\Omega^j_a F^j_b\partial_\omega p^j_b}} \exp\left[-i\int^{r_b}_{r_a}(1+i\Gamma\partial_\omega)p^j dr\right], \\
    R^j_a = & \ \mathcal{F}^j_{ab} R^j_b,
\end{split}
\end{equation}
and a single subscript indicates the location where a quantity is evaluated e.g.\ $R^j_a = R^j(r_a)$.
We then combine \eqref{tp1}, \eqref{tp2} and \eqref{shift} to define a relation between propagating amplitudes either side of the tunnelling zone $r_1<r<r_2$,
\begin{equation} \label{refl_formula}
\begin{split}
    \begin{pmatrix}
        R^+_1 \\ R^-_1
    \end{pmatrix} = & \  \mathcal{N}_{12}\begin{pmatrix}
        R^+_2 \\ R^-_2
    \end{pmatrix}, \\
    \mathcal{N}_{12} = & \ \mathcal{F}^\downarrow_{12}\begin{bmatrix}
        1+f_{12}^2/4 & i(1-f_{12}^2/4) \\ -i(1-f_{12}^2/4) & 1+f_{12}^2/4
    \end{bmatrix},
\end{split}
\end{equation}
where,
\begin{equation} \label{f12}
    f_{12} = \exp\left\{-\int^{r_2}_{r_1}\mathrm{Im}[(1+i\Gamma\partial_\omega)p^\downarrow]dr\right\}.
\end{equation}
For a wide tunnelling zone, $f_{12}$ is assumed small and can be evaluated to leading order for $\Gamma=0$.
These relations can now be used to find a resonance condition for modes that cross the tunnelling zone, by transporting the solution from $r_0$ in the vortex core (see Fig.~\ref{fig:4}) to the outer boundary at $r_B$,
\begin{equation} \label{drag}
\begin{split}
    R^+_0 = & \ \mathcal{F}^+_{01}\mathcal{F}^\downarrow_{12}\left(\sigma_+ \mathcal{F}^+_{2B}R^+_B + i\sigma_- \mathcal{F}^-_{2B}R^-_B \right), \\
    R^-_0 = & \ \mathcal{F}^-_{01}\mathcal{F}^\downarrow_{12}\left(-i\sigma_- \mathcal{F}^+_{2B}R^+_B + \sigma_+ \mathcal{F}^-_{2B}R^-_B \right), \\
    \sigma_\pm = & \ 1\pm f_{12}^2/4.
 \end{split}
\end{equation}
The final step is to impose the boundary conditions.
In the vortex core, regularity of the solution requires us to have $R^-_0 = iR^+_0$.
At the outer boundary, we can assume that the density goes to a constant value and that the velocities are vanishingly small provided we impose a Neumann boundary condition there; that is $R^+_B = R^-_B$.
Both of this steps are explained in detail in \cite{patrick2022quantum}.
Inserting these relations into \eqref{drag} and rearranging yields the resonance condition \eqref{cotcotexp} in the main text.
Note, the assumption of vanishingly small $v_r(r_B)$ means that square root factors in \eqref{shift} cancel on the boundary when computing $\mathcal{F}^+_{2B}/\mathcal{F}^-_{2B}$.

\end{document}